\documentclass[review]{elsarticle}
\usepackage{graphicx}
\usepackage{amsfonts,amssymb,amsmath}
\usepackage{bm}
\begin{document}
\begin{frontmatter}
\title{Influence of strain on the properties of CeRuPO and CeOsPO Kondo
systems}
\author{Matej Komelj}
\ead{matej.komelj@ijs.si}
\address{Jozef Stefan Institute, Jamova 39, SI-1000 Ljubljana, Slovenia}
\begin{abstract}
We investigated the influence of isotropic strain on the type of the
magnetic ground-state and the Fermi surface for the structurally-equivalent 
CeRuPO and CeOsPO 
crystals with slightly different lattice parameters. The two systems exhibit 
different magnetic orderings at zero strain.
According to the phase diagram of CeRuPO under pressure the difference
might be due to the different Ce-Ce 
inter-atomic distances.
We applied {\it ab-initio} calculations based on the 
density-functional
theory and the generalized-gradient approximation with an additional 
Coulomb repulsion (GGA+$U$), which indeed reveal a significant impact of
the strain and the  effective $U$ parameter on the magnetic ground state.
However,  it is demonstrated that the difference is more likely 
related to the details in the Fermi surface.
\end{abstract}
\end{frontmatter}
\section{Introduction}
The existence of different phases under the same conditions in otherwise 
similar systems represents one of the important themes in contemporary
condensed-matter physics\cite{Sachdev:2012}. Details in the electronic 
structure due
to slightly different chemical compositions and, consequently, a subtle 
variation of the lattice parameters may lead to 
substantially different ground states. A classical example is the 
competition between the ferromagnetic (FM) and the antiferromagnetic (AFM) 
orderings,
which is ascribed to different types of exchange coupling as well
as to the fine tuning of the corresponding parameters, like the inter-atomic
distances\cite{Radousky:2000}. Although the heavy-fermion systems are more often regarded as
antiferromagnetic, ferromagnetism in these materials is not so rare. 
Interestingly, most of the examples contain cerium as the source of 
magnetism\cite{Sullow:1999,Sidorov:2003,Larrea:2005,Drotziger:2006,Krellner:2007,
Das:2014,Jang:2014}. A special consideration is required for $\textrm{CeRuPO}$,
whose
isostructural sister $\textrm{CeOsPO}$ is antiferromagnetic\cite{Krellner:2007}.
Both, the ruthenium
and osmium compounds crystallize in the tetragonal $\textrm{ZrCuSiAs}$-type 
structure (space group $P4/nmm$) consisting of alternating $\textrm{RuP}_4$ 
($\textrm{OsP}_4$) and $\textrm{OCe}_4$-tetrahedra layers. The latter, which 
are magnetic, are well separated from each other. Therefore, a two-dimensional
nature of the properties is expected.  The unit cell of $\textrm{CeRuPO}$
determined from the experimental lattice parameters $a=4.028(1)\>\textrm{\AA}$  and
$c=8.256(2)\>\textrm{\AA}$) is only very slightly smaller than the unit cell of 
$\textrm{CeOSPO}$ with $a=4.031(1)\>\textrm{\AA}$ and 
$c=8.286(3)\>\textrm{\AA}$. Nevertheless, the 
two materials exhibit different magnetic ground states.
Recently, a pressure-dependent magnetic phase diagram for $\textrm{CeRuPO}$
was determined\cite{Kitagawa:2014}. The measured ${^{31}}$P-NMR 
spectra were explained by a FM-to-AFM phase transition, which should occur
when the sample is exposed to a pressure of about $0.7\>\textrm{GPa}$ at 
nearly zero temperature. Above $2.97\>\textrm{GPa}$ the measured signal indicated
the existence of a paramagnetic ground state.  Although the experiment 
covers just the unit-cell contraction due to the applied pressure, it is 
clear that the type of magnetic ordering can be determined by the strain
in the material. 
\par
In order to better understand the influence of strain and to explore its 
importance with respect to the difference between the properties of 
$\textrm{CeRuPO}$ and $\textrm{CeOsPO}$ we carried out a theoretical 
investigation
based on the density-functional theory (DFT). In fact, the corresponding band
structures obtained from a DFT calculation for the non-strained states were 
already discussed\cite{Krellner:2007}. The calculated 
total energies yielded the correct magnetic ground states. These band
structures revealed only subtle differences, among which the slightly different
oxygen hybridizations in both compounds were suggested as the most-likely 
reason for the different behavior of the two compounds. 
\section{Methods}
We applied the Quantum Espresso\cite{Giannozzi:2009} code and the 
generalized gradient approximation
(GGA)\cite{Perdew:1996}
for the exchange-correlation potential.  The electron-ion interactions
were described with Troullier-Martins-type\cite{Troullier:1991,Fuchs:1999} 
pseudopotentials. The strong correlations between the $4f$ Ce electrons were
treated by means of the simplified rotational-invariant GGA$+U$ 
scheme\cite{Cococcioni:2005}. Since the calculated properties, to some
extent, depend on the choice of the effective $U$ parameter (see, for example,
Ref. \cite{Loschen:2007}), we performed the calculations for different
values of $U=0$ (pure GGA), $2$, $4$ and $6\>\textrm{eV}$. 
On the basis of convergency tests the plane-wave and the charge-density 
cut-off parameters were  set to
$1020\>\textrm{eV}$ and $4080\>\textrm{eV}$, respectively, whereas a 
$4\times 4\times 2$ mesh of $\bf k$-points\cite{Monkhorst:1976} was used 
for the Brillouin-zone integration\cite{Blochl:1994}.
The criterion for the self-consistency was the total-energy difference
between the two subsequent iterations being less than $10^{-5}\>\textrm{Ry}$.
The theoretical equilibrium lattice parameters and the atomic positions were
determined by means of minimizing the total energies
and inter-atomic forces without taking into account spin the polarization and 
by setting the Coulomb repulsion to zero $U=0\>\textrm{eV}$.
The resulting $c/a$ ratio of $1.98$ ($1.97$) for $\textrm{CeRuPo}$ ($\textrm{CeOsPO}$)
differs from the experimental value of $2.05$ ($2.06$). 
The optimized structures were applied in the spin-polarized calculations 
for different types of magnetic ordering and values of $U$ by varying
the lattice parameter $a$, whereas the $c/a$ value was fixed. A decrease in $a$
has the same effect on the unit-cell volume as the application of an external 
pressure. In addition to the FM ordering
we considered only the most simple AFM arrangement of the Ce magnetic moments
pointing up and down at the two crystallographic sites within the unit cell.
\section{Results and discussion}
The total energies as a function of $a$ are presented in Fig. 1. 
The calculated values are fitted with third-order polynomials 
$f^U(a)=f^U_3a^3+f^U_2a^2+f^U_1a+f^U_0$ and 
$g^U(a)=g^U_3a^3+g^U_2a^2+g^U_1a+g^U_0$ for the FM and AFM states, respectively.
It is clear that the theoretical  equilibrium lattice parameter is
almost independent  of the type of magnetic ordering and grows very slightly
with an increasing $U$. The experimental value is exceeded by  
2.7\%-3.5\% in the case of the Ru and by 3.5\%-4.2\% in the case of the 
Os compound. 
Such a trend is in agreement with the GGA$+U$ results for cerium oxides from
Ref. \cite{Loschen:2007}.
\begin{figure}
\includegraphics[width=.8\textwidth]{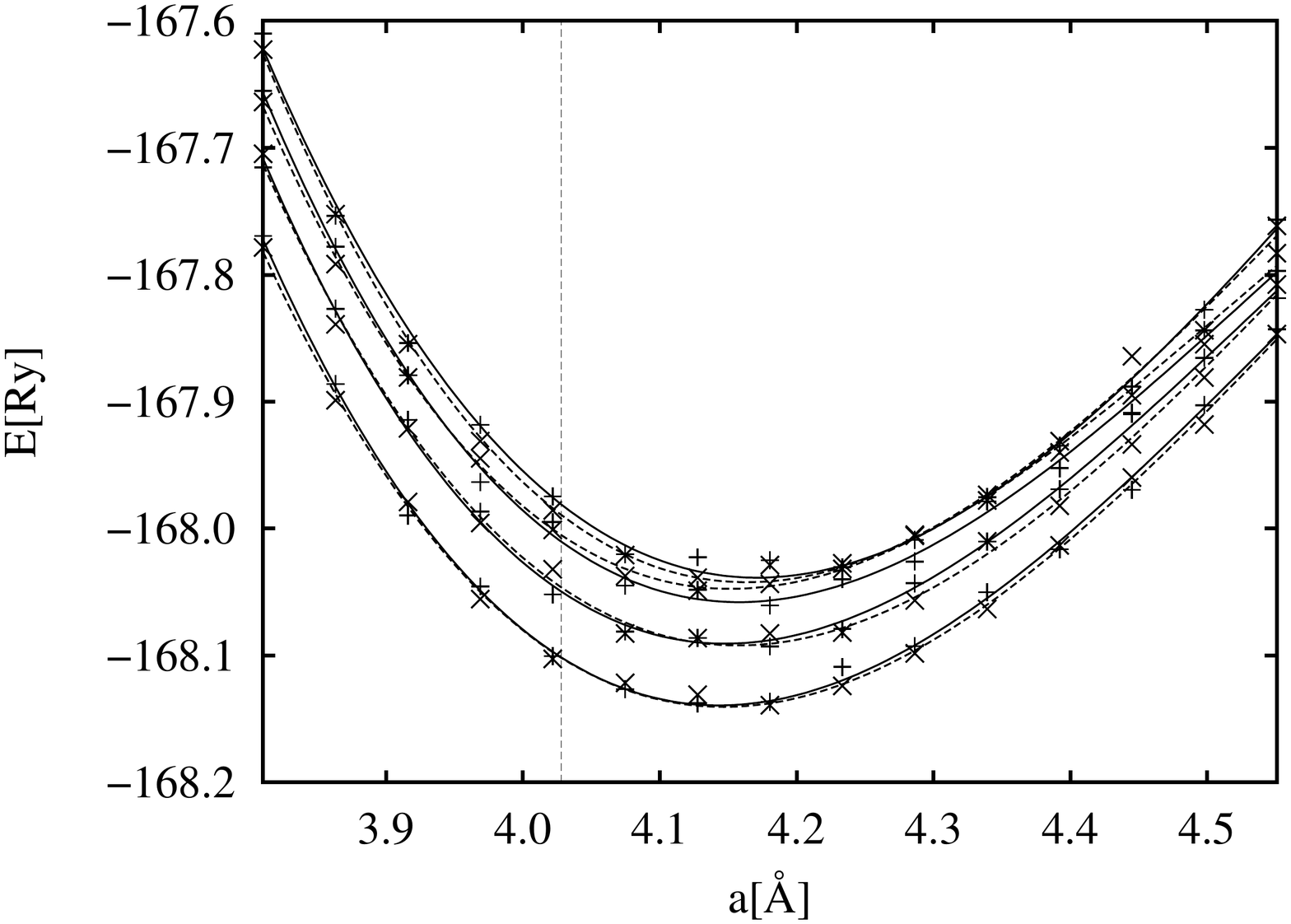}\\
\includegraphics[width=.8\textwidth]{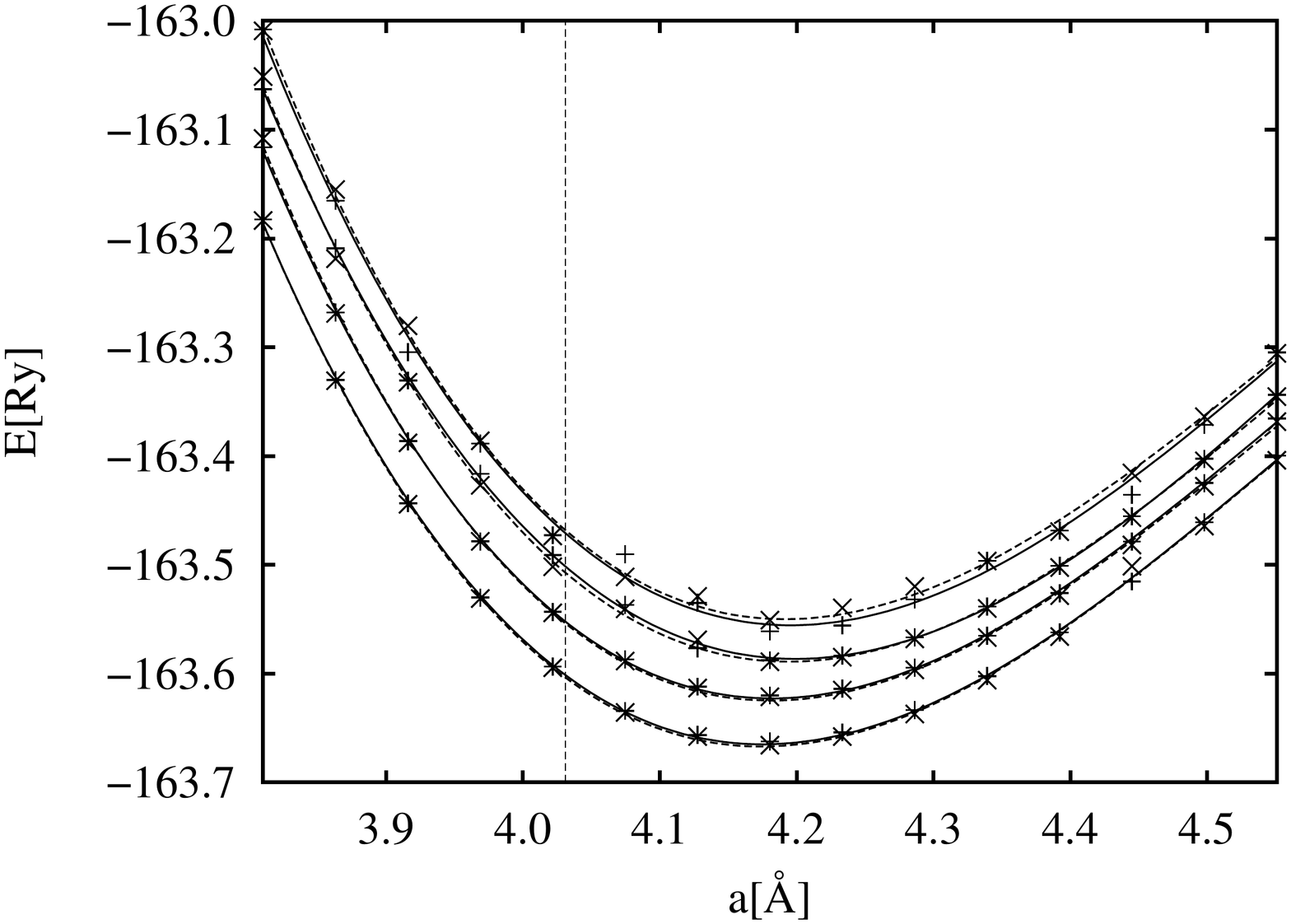}\\
\caption{The calculated values of the total energies 
for the FM ($+$) and AFM ($\times$) ordering in $\textrm{CeRuPO}$ (top)
and $\textrm{CeOsPO}$ (bottom). The data are fitted  with
third-order polynomials $f^U(a)$ for FM (solid lines) and $g^U(a)$ for
AFM ordering (dashed lines). The bunch of curves are for 
$U=0,2,4$ and $6\>\textrm{eV}$ from the bottom to the top. The vertical
lines are positioned at the values of the experimental lattice parameters.}
\end{figure}
The calculated values of the FM and AFM energies are very close to each
other.  Therefore,
it is more illustrative to examine the differences $f^U(a)-g^U(a)$ 
presented in Fig. 2. A positive value implies the AFM, and a negative value
implies the AFM ground state. The $a$ dependence and the influence of 
the $U$ value are pronounced. In the case of $\textrm{CeRuPO}$ only $U=4\>\textrm{eV}$
yields the proper FM ground state at the theoretical lattice parameter. 
For $U=2\>\textrm{eV}$ the FM ground state exists at the experimental 
lattice parameter, whereas for $U=6\>\textrm{eV}$ the FM state is energetically
more favorable at larger values of $a$. At a strain of $\sim 8.4\%$ 
the results of the calculation with $U=4\>\textrm{V}$ predict a transition 
to the AFM state, which is in qualitative agreement with the 
experiment\cite{Kitagawa:2014}. Krellner {\it et al.}\cite{Krellner:2007} 
applied the LSDA$+U$ method with $U=6.4\>\textrm{eV}$ and found the correct
FM ground state, but at the experimental lattice parameter. The system
$\textrm{CeOsPO}$ is less sensitive to the choice of U. Already a pure GGA 
calculation with $U=0\>\textrm{eV}$ leads to the proper AFM ground state,
which is also the case for $U=2\>\textrm{eV}$ and $U=4\>\textrm{eV}$, whereas
$U=6\>\textrm{eV}$ predicts the FM ordering as being preferential for the whole
range of $a$. Since there are no experimental results available for 
a strained $\textrm{CeOsPO}$ it is hard to determine which of the $U$ 
values is more appropriate. Both values $U=2$ and $4\>\textrm{eV}$ predict 
a transition to the FM state at high strains, which might be proved
experimentally. In addition, for $U=4\>\textrm{eV}$ the FM state 
prevails at a modest expansion of the unit cell. 
\begin{figure}
\includegraphics[width=.8\textwidth]{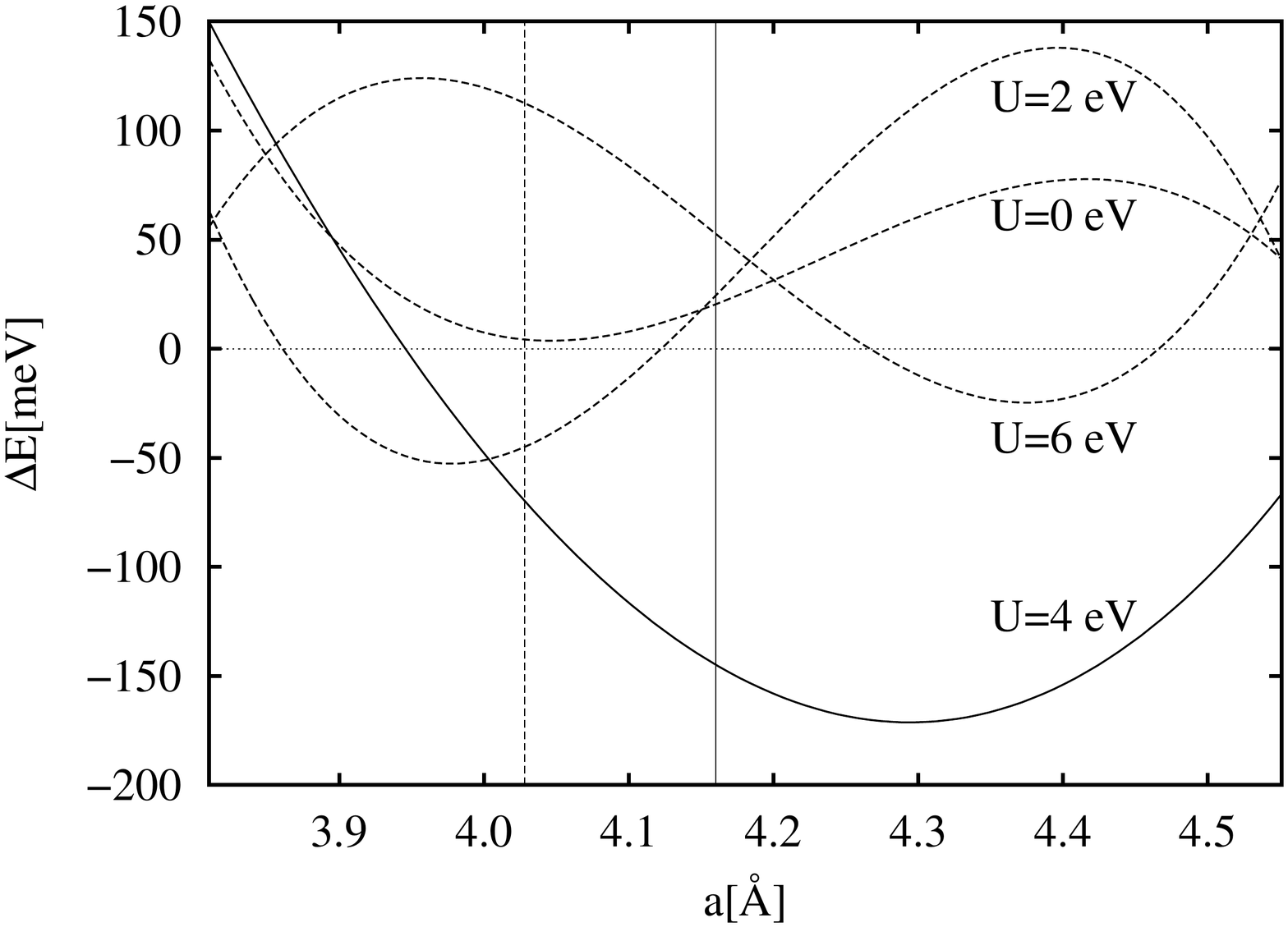}\\
\includegraphics[width=.8\textwidth]{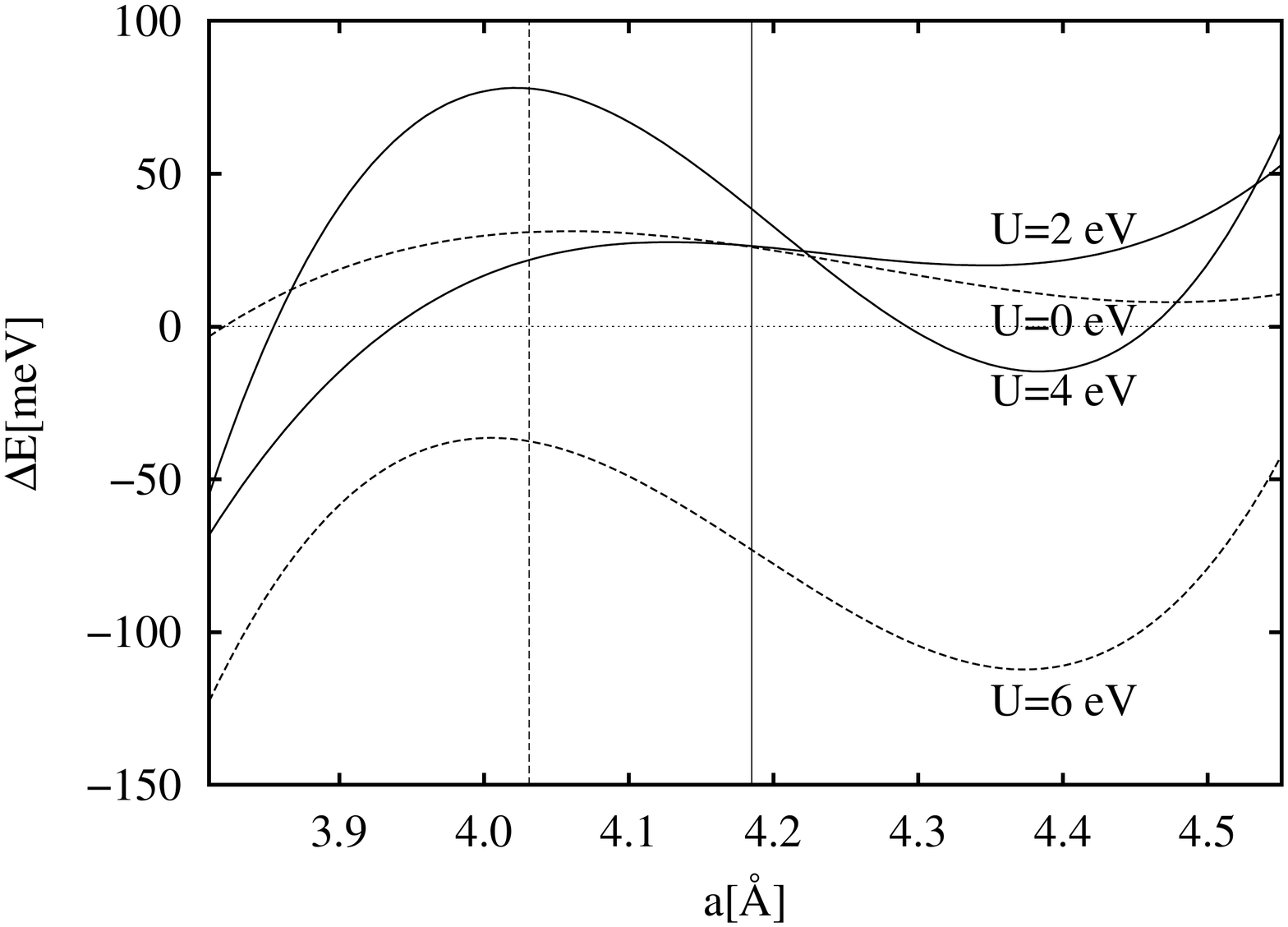}\\
\caption{
The difference in the total energy between the FM and the AFM
states for $\textrm{CeRuPO}$ (top) and $\textrm{CeOsPO}$ (bottom) obtained
from the  3rd-order-polynomial fits to the calculated values as a function 
of the lattice parameter $a$ for different $U$ values. The dashed vertical 
line is at the experimental and the solid vertical line is at the theoretical 
value of the lattice parameter. Positive values imply the AFM and negative
values the FM ground state. Solid curves represent the theoretical 
predictions that reproduce the experimentally-observed behavior.}
\end{figure}
The calculated behavior of $\textrm{CeOsPO}$ in principle 
supports the idea of the lattice-parameter mismatch as being the driving force
for the difference in the magnetic ground states of the Ru and Os compounds. 
But the values of $a$ at which the  FM state becomes energetically favorable
are much smaller than the equilibrium lattice parameter of $\textrm{CeRuPO}$. 
Furthermore, the calculations for $\textrm{CeRuPO}$ do not exhibit
the opposite behavior: within the whole considered range of the positive 
strain the FM ordering remains in the ground state. \par
As demonstrated by Yamamoto and Si\cite{Yamamoto:2010} the topology of the
Fermi surface (FS) is one of the crucial factors for the type of magnetism in
heavy-fermion systems. The calculated FS for $U=4\>\textrm{eV}$ 
and different values of $a$ are presented in Fig. 3. At zero strain
both, $\textrm{CeRuPO}$ and $\textrm{CeOsPO}$ FSs consist of three sheets 
per spin channel. As mentioned by Krellner {\it et al.}\cite{Krellner:2007}
the latter is represented by nearly perfect 
cylindrical tubes, typical for layered systems that exhibit two 
dimensionality\cite{Lebegue:2007}. In the case of the former, the inner-most tubes are 
deformed into pairs of mirror-symmetrical cones, which hints a 
non-negligible interaction between the $\textrm{RuP}_4$ and 
$\textrm{OCe}_4$-tetrahedra layers. This hypothesis is supported by the absence 
of cones in the FS for $\textrm{CeRuPO}$ at large values of the lattice parameter
$a=8.6\>\textrm{\AA}$. Since the $c/a$ ratio is kept fixed the inter-layer 
distance grows too, and consequently the corresponding interaction gets
weaker. The $\textrm{CeOsPO}$ FS for the same value of the lattice parameter
is even more cylindrical than in the case of the non-strained state, and
all three sheets per spin are preserved although the inner-most tubes become
very narrow. The outer-most tubes in the FS's of $\textrm{CeRuPO}$ and 
$\textrm{CeOsPO}$ for $a=7.2\>\textrm{\AA}$ are similarly distorted, indicating
a more pronounced inter-layer interactions. The cones in the case of
the Ru compound are even more significant than they are for the non-strained
state. In the case of the Os compound the cylindrical tubes are almost
non-distorted, although a pair of tiny cones appears for the down-spin
channel as an additional fourth sheet. 
\begin{figure*}
\includegraphics[angle=-90,width=1.\textwidth]{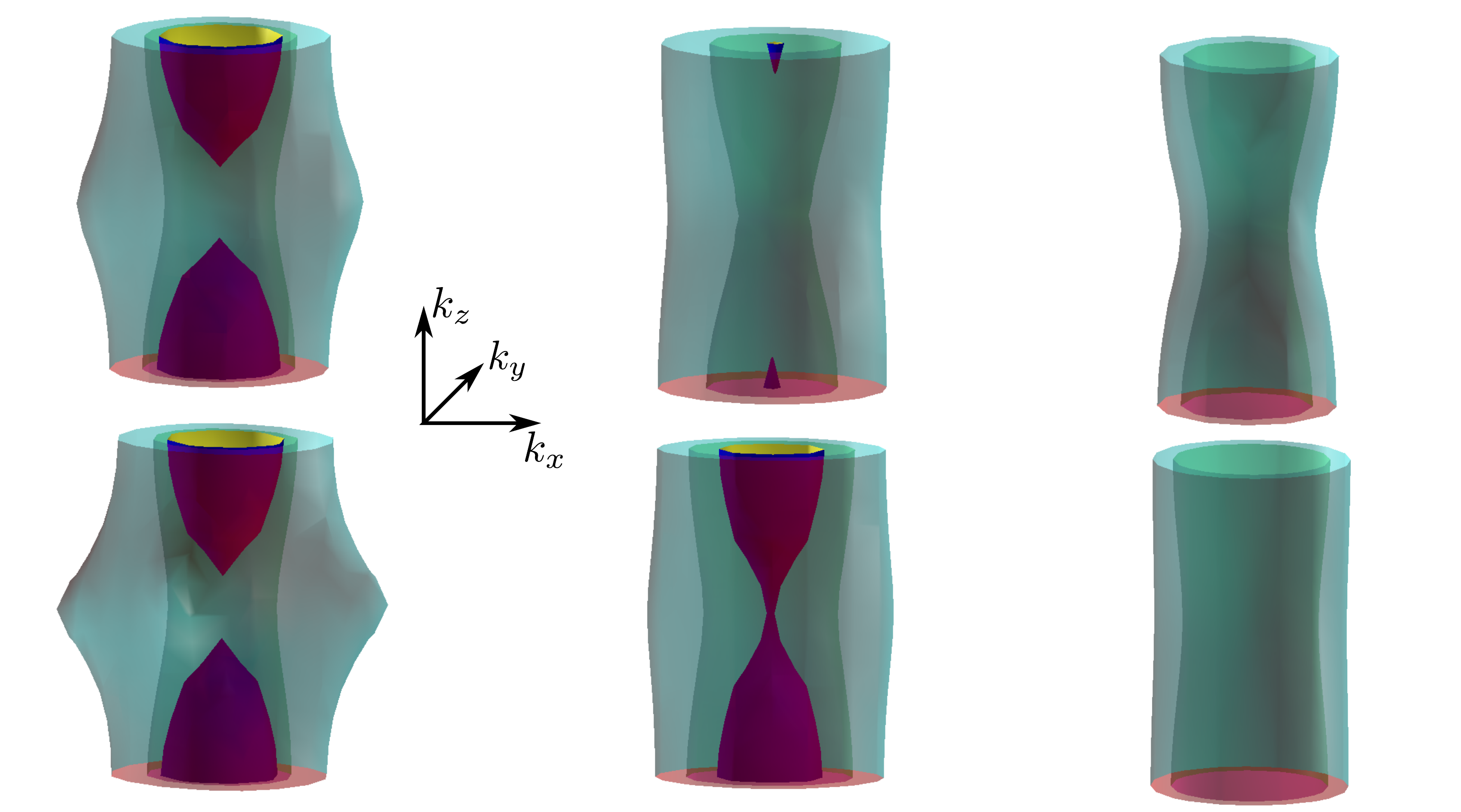}\\
\includegraphics[angle=-90,width=1.\textwidth]{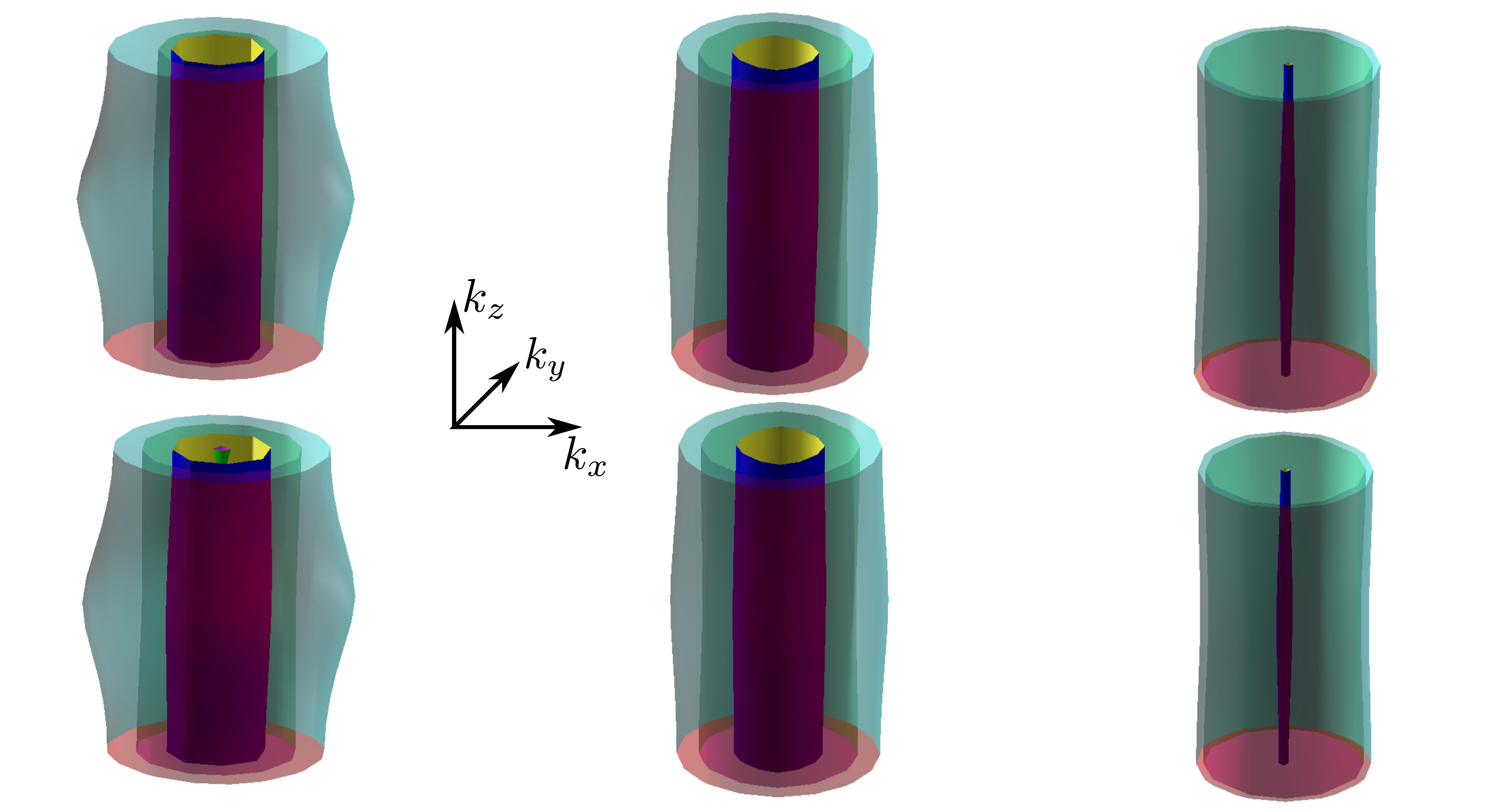}
\caption{(Color online) The calculated Fermi surfaces for $\textrm{CeRuPO}$ (top) 
and $\textrm{CeOsPo}$ (bottom) by applying $U=4\>\textrm{eV}$.  In each figure the 
first row is for the 
spin up, and the second row is for the spin down. The left
column is for $a=7.2\>\textrm{\AA}$, the middle column is for the 
theoretical equilibrium lattice parameter, and the right column
is for $a=8.6\>\textrm{\AA}$.}
\end{figure*}
\section{Conclusion}
On the basis of {\it ab-initio} calculations we determined
the magnetic ground states for $\textrm{CeRuPO}$ and $\textrm{CeOsPO}$ 
isostructural systems as a function of the lattice parameter and the 
Coulomb-repulsion parameter $U$ for the Ce $4f$ electrons. For $U=4\>{\rm eV}$
all the available experimental results were fairly reproduced. The
calculated Fermi surface of $\textrm{CeRuPO}$ exhibits a less pronounced 
two-dimensional character than the one of $\textrm{CeOsPO}$ due to a stronger
interaction between the layers within the crystal structure, which is 
probably the main reason for the different magnetic ground states of the 
two materials. \par
We call for a measurement of the pressure-dependent $\textrm{CeOsPO}$ 
magnetic phase diagram.
\bibliography{text.bib}

\begin{thebibliography}{10}
\expandafter\ifx\csname url\endcsname\relax
  \def\url#1{\texttt{#1}}\fi
\expandafter\ifx\csname urlprefix\endcsname\relax\def\urlprefix{URL }\fi
\expandafter\ifx\csname href\endcsname\relax
  \def\href#1#2{#2} \def\path#1{#1}\fi

\bibitem{Sachdev:2012}
S.~Sachdev, arXiv:1203.4565v4.

\bibitem{Radousky:2000}
H.~B. Radousky, Magnetism in Heavy Fermion Systems, World Scientific Publishing
  Co. Pte. Ltd., PO Box 128, Farrer Road, Singapore 912805, 2000.

\bibitem{Sullow:1999}
S.~S\"{u}llow, M.~C. Aronson, B.~D. Rainford, P.~Haen, Phys. Rev. Lett. 82
  (1999) 2963.

\bibitem{Sidorov:2003}
V.~A. Sidorov, E.~D. Bauer, N.~A. Frederick, J.~R. Jeffries, S.~Nakatsuji,
  N.~O. Moreno, J.~D. Thompson, M.~B. Maple, Z.~Fisk, Phys. Rev. B 67 (2003)
  224419.

\bibitem{Larrea:2005}
J.~{Larrea J.}, M.~B. Fontes, A.~D. Alvarenga, E.~M. Baggio-Saitovitch,
  T.~Burghardt, A.~Eichler, M.~A. Continentino, Phys. Rev. B 72 (2005) 035129.

\bibitem{Drotziger:2006}
S.~Drotziger, C.~Pflelderer, M.~Uhlarz, H.~v.~L\"{o}hneysen, D.~Souptel,
  W.~L\"{oser}, G.~Behr, Phys. Rev. B 73 (2006) 214413.

\bibitem{Krellner:2007}
C.~Krellner, N.~S. Kini, E.~M. Br\"{u}ning, K.~Koch, H.~Rosner, M.~Nicklas,
  M.~Baenitz, C.~Geibel, Phys. Rev. B 76 (2007) 104418.

\bibitem{Das:2014}
D.~Das, T.~Gruner, H.~Pfau, U.~B. Paramanik, C.~Geibel, Z.~Hossain, J. Phys.
  Condens. Matter 26 (2014) 106001.

\bibitem{Jang:2014}
H.~Jang, G.~Friemel, J.~Ollivier, A.~V. Dukhnenko, N.~Y. Shitsevalova, V.~B.
  Filipov, B.~Keimer, D.~S. Inosov, Nature Mater. 13 (2014) 682.

\bibitem{Kitagawa:2014}
S.~Kitagawa, H.~Kotegawa, H.~Tou, R.~Yamauchi, E.~Matsuoka, H.~Sugawara, Phys.
  Rev. B 90 (2014) 134406.

\bibitem{Giannozzi:2009}
P.~Giannozzi, S.~Baroni, N.~Bonini, M.~Calandra, R.~Car, C.~Cavazzoni,
  D.~Ceresoli, G.~L. Chiarottia, M.~Cococcioni, I.~Dabo, A.~D. Corso,
  S.~Fabris, G.~Fratesi, S.~de~Gironcoli, R.~Gebauer, U.~Gerstmann,
  C.~Gougoussis, A.~Kokalj, M.~Lazzeri, L.~Martin-Samos, N.~Marzari, F.~Mauri,
  R.~Mazzarello, S.~Paolini, A.~Pasquarello, L.~Paulatto, C.~Sbraccia,
  S.~Scandolo, G.~Sclauzero, A.~P. Seitsonena, A.~Smogunov, P.~Umari, R.~M.
  Wentzcovitch, J. Phys. Condens. Matter 21 (2009) 395502.

\bibitem{Perdew:1996}
J.~P. Perdew, K.~Burke, M.~Ernzerhof, Phys. Rev. Lett. 77 (1996) 3865.

\bibitem{Troullier:1991}
N.~Troullier, J.~L. Martins, Phys. Rev. B 43 (1991) 1993.

\bibitem{Fuchs:1999}
M.~Fuchs, M.~Scheffler, Comput. Phys. Commun. 119 (1999) 67.

\bibitem{Cococcioni:2005}
M.~Cococcioni, S.~de~Gironcoli, Phys. Rev. B 71 (2005) 035105.

\bibitem{Loschen:2007}
C.~Loschen, J.~Carrasco, K.~M. Neyman, F.~Illas, Phys. Rev. B 75 (2007) 035115.

\bibitem{Monkhorst:1976}
H.~J. Monkhorst, J.~Pack, Phys. Rev. B 13 (1976) 5188.

\bibitem{Blochl:1994}
P.~E. Bl\"{o}chl, O.~Jepsen, O.~K. Andersen, Phys. Rev. B 49 (1994) 16223.

\bibitem{Yamamoto:2010}
S.~J. Yamamoto, Q.~Si, Proc. Natl. Acad. Sci. U.S.A. 107 (2010) 15704.

\bibitem{Lebegue:2007}
S.~Leb\'{e}gue, Phys. Rev. B 75 (2007) 035110.

\end{thebibliography}
\end{document}